\documentclass[twocolumn,trackchanges]{aastex63}

\received{ TBD }       
\revised{\today}      
\accepted{TBD } 

\shorttitle{Study of radio transients from the quiet Sun during an extremely quiet time}
\shortauthors{Mondal et al.}

\begin{document}

\title{Study of radio transients from the quiet Sun during an extremely quiet time}

\correspondingauthor{Surajit Mondal}
\email{surajit.mondal@njit.edu}

\author[0000-0002-2325-5298]{Surajit Mondal}
\affiliation{Center for Solar Terrestrial Research, New Jersey Institute of Technology, Newark, New Jersey, 07102, United States of America}

\author[0000-0002-4768-9058]{Divya Oberoi}
\affiliation{National Centre for Radio Astrophysics, Tata Institute of Fundamental Research, Pune-411007, India}

\author[0000-0002-1741-6286]{Ayan Biswas}
\affiliation{National Centre for Radio Astrophysics, Tata Institute of Fundamental Research, Pune-411007, India}

\begin{abstract}

In this work we study a class of recently discovered metrewave solar transients referred to as Weak Impulsive Narrowband Quiet Sun Emission \citep[WINQSEs,][]{mondal2020}.
Their strength is a few percent of the quiet Sun background and are characterised by their very impulsive, narrow-band and ubiquitous presence in quiet Sun regions. 
\citet{mondal2020} hypothesised that these emissions might be the radio counterparts of the nanoflares and their potential significance warrants detailed studies.
Here we present an analysis of data from an extremely quiet time and with an improved methodology over the previous work.
As before, we detect numerous WINQSEs, which we have used for their further characterisation. 
Their key properties, namely, their impulsive nature and ubiquitous presence on the quiet Sun are observed in these data as well.
Interestingly, we also find some of the observed properties to differ significantly from the earlier work.
With this demonstration of routine detection of WINQSEs, we hope to engender interest in the larger community to build a deeper understanding of WINQSEs.

\end{abstract}

\keywords{\href{http://astrothesaurus.org/uat/1992}{Quiet solar corona (1992)}; \href{http://astrothesaurus.org/uat/1483}{Solar corona (1483)}; \href{http://astrothesaurus.org/uat/1993}{Solar coronal radio
emission (1993)}; \href{http://astrothesaurus.org/uat/1989}{Solar coronal heating (1989)}}

\section{Introduction}\label{sec:intro}

Radio transients from the Sun were discovered at very the dawn of solar radio astronomy \citep[e.g.][etc.]{wild1963, wild1972}.
Since then numerous studies have analysed and studied their various properties ranging from emission mechanisms, spectro-temporal characteristics, relationships with other solar phenomenon, signatures in other wavebands and so on \citep[e.g.][and many others]{takakura1967,sturrock1964,melrose1980,mclean1985,pick2008,kontar2017,reid2017,reid2020}. 
Most of these transients, particularly those happening in the meter wavelengths regime, are associated with active regions and violent solar activity like X-ray flares and coronal mass ejections \citep[e.g.][etc.]{pick2008,bain2009,krucker2011,reid2020,mondal2021a}. 
However, there are hypotheses which suggest that such radio transients, albeit much weaker, should also be present in the quiet solar corona \citep{che2018}. The primary reasoning behind this expectation comes from the Parker's nanoflare hypothesis \citep{parker1988} which states that the solar corona maintains its million K temperature because of the energy continuously deposited by numerous tiny nanoflares happening everywhere in the corona. 
If this hypothesis is true, it implies that small scale reconnections are happening everywhere and all the time in the corona. 
These small scale reconnections, like their larger counterparts, will also accelerate electrons, which in turn, on their interaction with the thermal plasma, should emit plasma emission, observable in the radio bands \citep{che2018}. 
To the best of our knowledge, the first detection of the ubiquitous presence of these radio transients from the quiet sun came from \citet{mondal2020} (henceforth referred to as M20). 
They found that these transients are highly impulsive (duration $\lesssim 1$s) and narrowband in nature. Based on these properties they also concluded that these emissions are generated by nonthermal processes. 
They also found these emissions to be present throughout the quiet sun.
 \citet{mondal2021b} has demonstrated, although in a single instance, that the energy associated with a group of these transients is $\sim 10^{25}$ergs. These observations suggest the possibility that these transients are the radio counterparts of the long hypothesised nanoflares. Henceforth we refer to these emissions as Weak Impulsive Narrowband Quiet Sun Emissions (WINQSEs). 

Due to the large potential significance of this discovery, it is crucial to verify the presence of WINQSEs in other datasets obtained under different solar conditions, using independent detection techniques and especially during quiet Sun times.
{ As a part of this exercise, recently \citet{sharma2022} (henceforth referred to as S22) have reported detection of WINQSEs using a completely independent technique and analysis pipeline using a different dataset.
Interestingly S22 also identified a very weak type III solar radio burst like drifting emission feature, which supports the hypothesis that WINQSEs are the weaker cousins of the stronger type III bursts.}
Continuing with the same motivation, here we present an analysis of data from an extremely quiet time on 20$^{th}$ June, 2020. The solar activity levels on this day are much lower than those prevailing during the observations of M20 and S22. We also take this opportunity to improve the analysis methodology beyond what was used in M20.

This paper is structured as follows. In Section 2 we present the observation details, Section 3 discusses the details of data analysis including the improvements over M20. In Section 4 we present the results, in Section 5 the discussion and finally Section 6 gives the concluding remarks.

\section{Observation details}

The data presented here come from a solar observation with the Murchison Widefield Array Phase II \citep[MWA,][]{lonsdale2009, tingay2013, wayth2018}.
These data were acquired on 20 June, 2020 between 03:35:00--04:40:00 UT under the project ID G0002. 
The MWA was in its extended configuration, with a maximum baseline of $\sim 5$km. The observing frequency spanned the band from 119.68--150.40 MHz. 
The data was acquired with a time and frequency resolution of 0.5 s and 10 kHz respectively. 
Here we present the analysis at four selected spectral slices, each of width 160 kHz.
These slices are centered around 120.22, 127.9, 135.58 and 139.9 MHz. 

The Sun was extremely quiet on this day\footnote{\url{https://solarmonitor.org/index.php?date=20200620}}. No X-ray flares were reported by the Geostationary Operational Environmental Satellites (GOES) and no events of any kind were reported by the Space Weather Prediction Center event list for the day. 
The Learmonth Spectrograph operating between 25--180 MHz also did not report any radio flare on this day. No NOAA active region was present on the visible solar disc.

\section{Data analysis}
\label{sec:analysis}

Making the large number of images needed for this investigation necessarily requires a robust automated 
pipeline. 
Such a tool has recently been developed and christened the Automated Imaging Routine for Compact Arrays for Radio Sun \citep[AIRCARS;][]{mondal2019}. 
The credentials of this self-calibration based pipeline have already been established in multiple recent works requiring high quality spectroscopic snapshot imaging \citep[e.g.][etc.]{mohan2019,mondal2020,mohan2021,mondal2021a}.
It is known from past experience that the presence of strong compact sources on the Sun makes calibration easier and the data corresponding to a large but featureless quiet Sun is harder to calibrate  \citep{mondal2019}.
While the imaging of the data used here was done using AIRCARS, to improve the accuracy of calibration for these quiet Sun data, an improved strategy from what was followed in M20 was employed. 
For this work, the calibration solutions were determined using data averaged over 9 s. 
This time integration helps reduce the thermal noise substantially and is especially useful in improving the signal-to-noise of visibilities from longer baselines.
The calibration solutions obtained were linearly interpolated in time and applied to the 0.5 s resolution data which were then imaged.
The analysis for each of the four frequencies presented here was carried out independently following this procedure.
This approach relies on the assumption that the solar variability over the integration period is small enough to be ignored. 
We find this to be a good assumption for this extremely quiet time and is evidenced by the higher imaging dynamic ranges of the images produced using solutions derived from time-averaged visibilities.

On its own, AIRCARS does not ensure a common flux density scale across independent observations.
For the present analysis, this can lead to discontinuities in the fluxscale at the boundaries of observations, typically of a duration of 4-5 minutes, though this issue has now been resolved in its next incarnation, named, Polarimetry using Automated Imaging Routine for Compact Arrays for the Radio Sun \citep[P-AIRCARS;][]{kansabanik2022a}.
A precise absolute flux density is not required for this work and we have adopted the strategy described below for relative flux density calibration between all of the observations at a given frequency. 
It also takes care of the amplitude variability of the order of 4-5\% of the MWA instrumental response which otherwise is not possible to take care of by using the method described in \citet{kansabanik2022b}. 
While under most circumstances, this level of variability is too small to be of any consequence, in the present context it is important to correct for it as it is of a magnitude comparable to the strengths of WINQSEs.

A correction for the instrumental primary beam is first applied to the solar maps and then it is enforced that the median disc integrated flux density of the Sun be the same for all five-minute chunks of data used.
The underlying assumption is that as the Sun was very quiet during our observing period, its disc-integrated flux density should also be stable in this period.
The characteristics of WINQSEs - flux densities of $\sim$ mSFU (as compared to a few SFU from the Sun) and short durations ($\lesssim$1 s) - imply that they are too weak to influence the disc-integrated flux density in any significant manner.
We also found that the point-spread-function (psf) corresponding to different time slices for a given frequency can change slightly across time.
 Flux densities in radio maps are measured in units of $Jy/beam$, where $beam$ refers to the area of the psf.
Hence, slight changes in the psf area can give rise to low-level jumps in flux density estimates.
To avoid contamination due to this, we have smoothed all of the images used here to a common 
coarser resolution of $280^{"}\times 280^{"}$.

\begin{figure}
    \centering
    \includegraphics[trim={0 0 0 0.5cm},clip,scale=0.5]{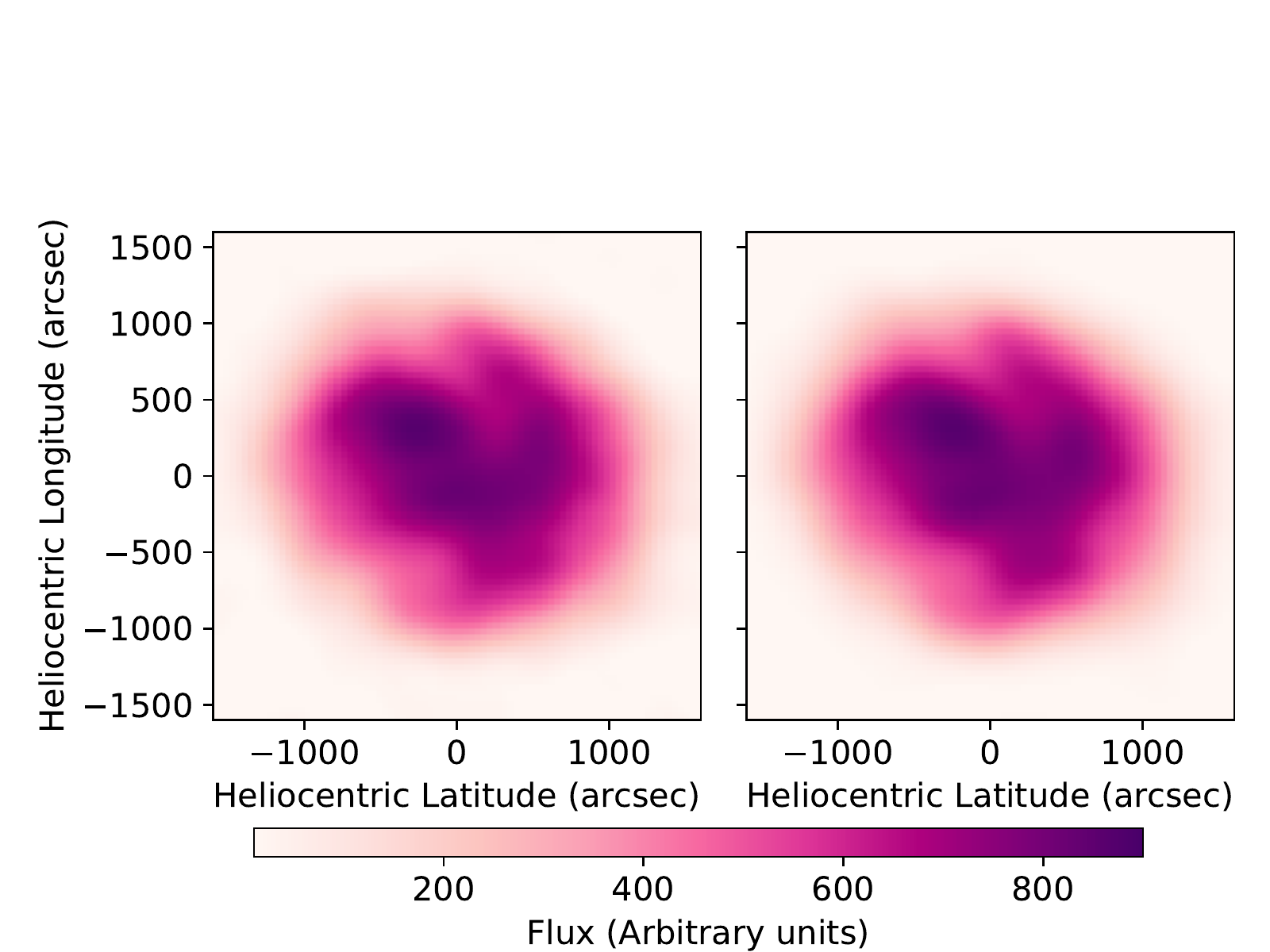}
    \caption{Left panel: An example 0.5 s map at 127.9 MHz. Right panel: The corresponding median map at 127.9 MHz.}
    \label{fig:sample_image_paper}
\end{figure}

A key step for identifying WINQSEs in M20 
was to calculate the median image, which requires accurate alignment of images relative to each other.
M20 achieved this by aligning the radio peak of the active region present in their dataset across images.
Absence of a clearly identifiable compact feature forces us to look for other alternatives to achieve this objective.
We compute a center of intensity ($\hat{X}$, $\hat{Y}$) for each 0.5 s solar image, defined as:

\begin{equation}
    \hat{X}=\frac{\sum_i x_i}{\sum_i 1}, \quad
    \hat{Y}=\frac{\sum_i y_i}{\sum_i 1},
\end{equation}
where ($x_i$, $y_i$) are the coordinates of the pixels which exceed the threshold chosen visually such that no noise contours appear in the image and the region it demarcates is reasonably symmetric.
The centers of intensity for all the images were aligned to lie at the same pixel.
The aligned maps were used to generate the median map, one for each five-minute data chunk and frequency. An example 0.5 s map at 127.9 MHz and the corresponding median map are shown in Fig. \ref{fig:sample_image_paper}.

\section{Results}

\begin{figure}
    \centering
    \includegraphics[scale=0.5]{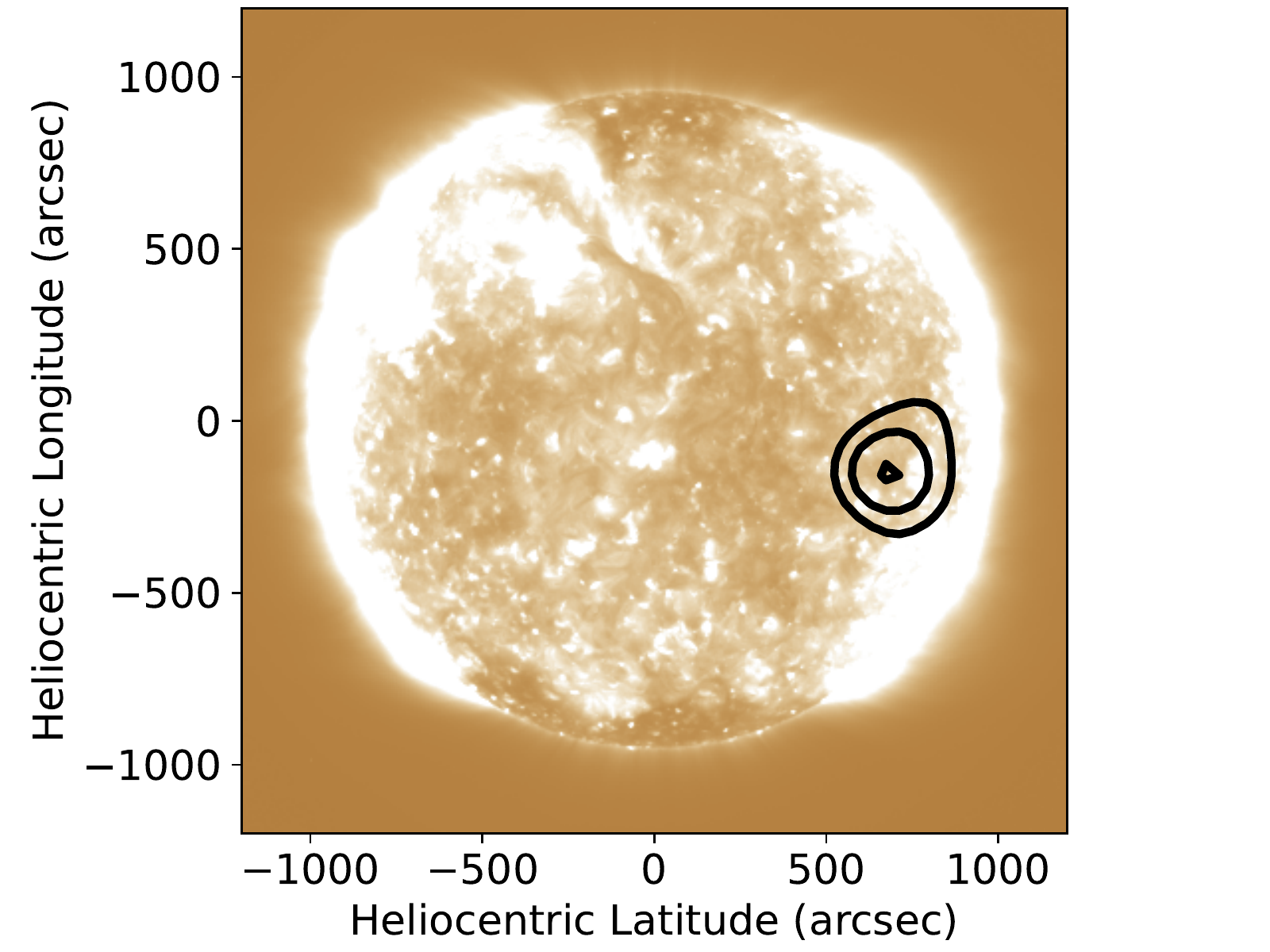}
    \caption{Radio contours of an example WINQSE at 136 MHz are overlaid on a 193 \AA$\,$ image from AIA/SDO. The contour levels are at 0.7, 0.85 and 0.99 times the peak of the median subtracted map. }
    \label{fig:WINQSE_structure}
\end{figure}

For detection of WINQSEs in the large number of images used in this work, we have modified the technique used in M20. First, all 0.5 s solar images were smoothed using a  circular kernel of size 280$^{"}$. Let the images prior and post this smoothing operation be denoted by $I_{init}$ and $I_{flux}$ respectively. 
As mentioned in the previous section, this smoothing operation ensured that the numerical value of each pixel in the radio image
$I_{flux}$ is equal to the flux density for the resolution element corresponding to the circular smoothing kernel centered on that pixel in $I_{init}$.  The median map obtained using $I_{init}$ images was used to obtain $I_{median,flux}$ using the same technique as that used to obtain $I_{flux}$. 
$I_{median,flux}$ was then subtracted from each $I_{flux}$ to obtain the residual maps for each 0.5 s timeslice, henceforth referred to as $I_{res}$. The pixel values in $I_{res}$ is equivalent to $\Delta F_{i,\nu,t} = F_{i,\nu,t}-<F_{i,\nu}>$, where $F_{i,\nu,t}$ is the flux density of a psf sized region $i$ at frequency $\nu$ some time $t$ and $<F_{i,\nu}>$ is the median flux density for region $i$ at frequency $\nu$. A WINQSE is said to be present when the pixel value in $I_{res}$ exceeds a given threshold. 
In this work, the threshold is set to $3\sigma$, where $\sigma$ is the rms in the corresponding $I_{flux}$ in a region away from the Sun. To ensure that only independent pixels are searched for WINQSEs, the search algorithm steps through the pixels in steps of 7 pixels in both directions which translates to $280^{"}$ in each direction and is equal to the effective image resolution.
Figure \ref{fig:WINQSE_structure} shows the contours of an example residual image with a bright WINQSE, overlaid on a AIA 193\AA$\,$ image.
The compact feature seen in the image is the WINQSE detected in this snapshot and it is evident that it has been detected with good signal-to-noise. The following subsections discuss the various properties of the WINQSEs detected in these data.

\subsection{Flux density distribution}
\begin{figure}
    \centering
    \includegraphics[scale=0.5]{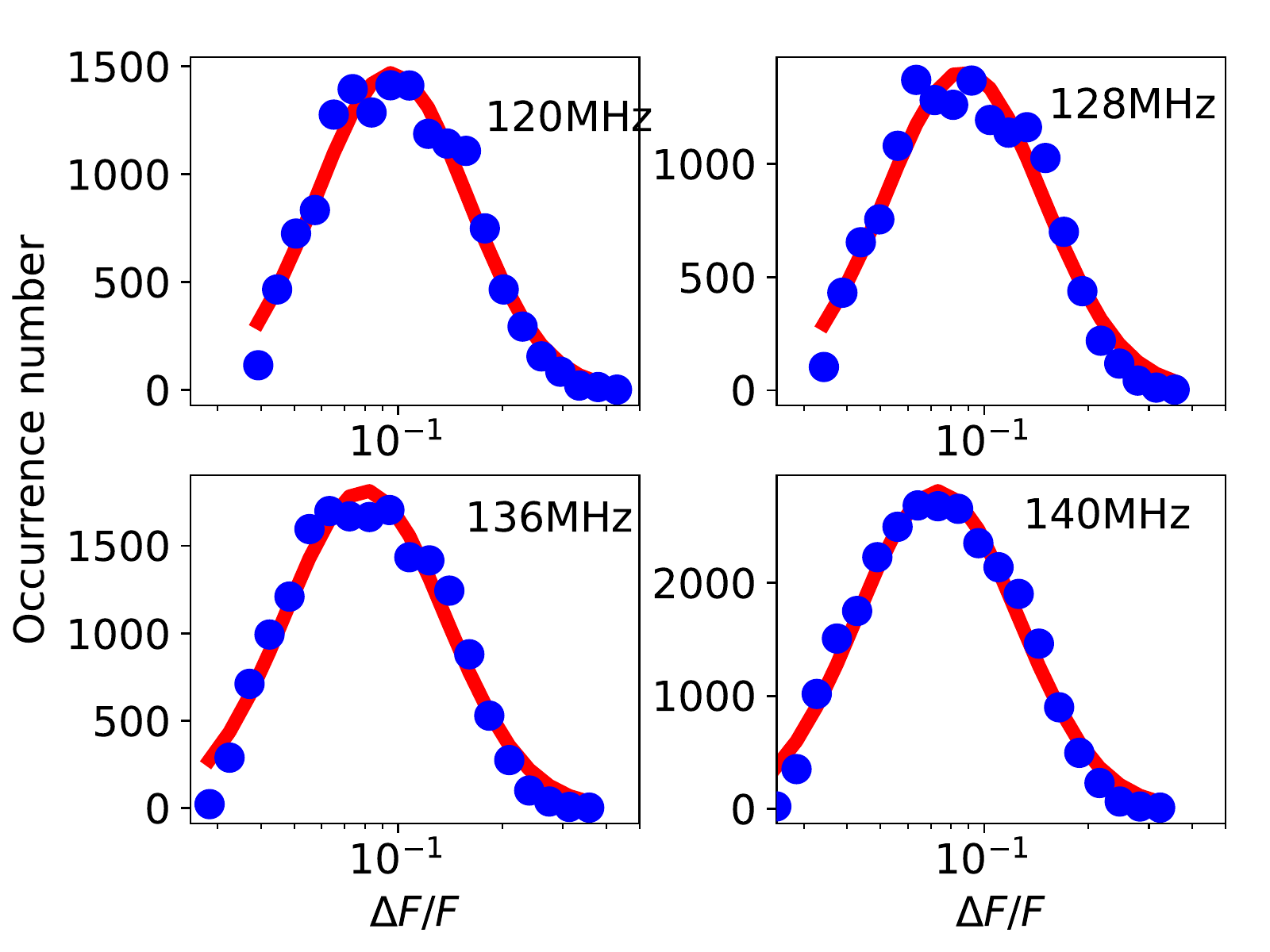}
    \caption{Histograms of $\Delta F/F$ are shown using the blue circles. The red line shows the lognormal fit to these data. The observation frequencies are mentioned in each panel.}
    \label{fig:deltaF_histogram}
\end{figure}

For convenience and consistency with M20, for a given frequency $\nu$, we refer to the collection of $\Delta F_{i,\nu,t}$ for all psf sized regions $i$ s and all times $t$ s as $\Delta F$ and similarly to $<F_{i,\nu}>$ as $F$. 
We define $\Delta F/F$ to be the ratio of the flux density in residual images ($I_{res}$) to that in the corresponding median image ($I_{median,flux}$) at the same pixel location.
Figure \ref{fig:deltaF_histogram} shows the histogram of $\Delta F/F$ in detected WINQSEs.
These distributions are strikingly different from those found by M20. 
In Sec. \ref{subsec:M20comarison} we try to understand and reconcile these differences.

The red line in Fig. \ref{fig:deltaF_histogram} shows a lognormal fit to the data shown with blue points. It is evident that the lognormal distribution is a good fit to the data. However, it is worth noting that the part of the distribution where $\Delta F/F \lesssim 0.1$ is affected by incompleteness due to a combination of multiple factors including sensitivity, and spatial and temporal resolution. These factors are not expected to significantly affect the data at higher $\Delta F/F$ values and hence this part of the histogram is robust. 
The fact that such lognormal distributions have been observed in earlier works when studying weak impulsive emissions, though at EUV wavelengths and with rather different temporal and spatial sampling \citep{pauluhn2007,tajfirouze2012,upendran2021}, suggests that WINQSEs too might follow a lognormal distribution. Testing this hypothesis in a more robust manner requires an instrument with better sensitivity and resolution. We plan to explore this with the next phase of the MWA (Phase III) which will offer a factor of two larger effective collecting area.

\subsection{Temporal width distribution}

\begin{figure}
    \centering
    \includegraphics[scale=0.5]{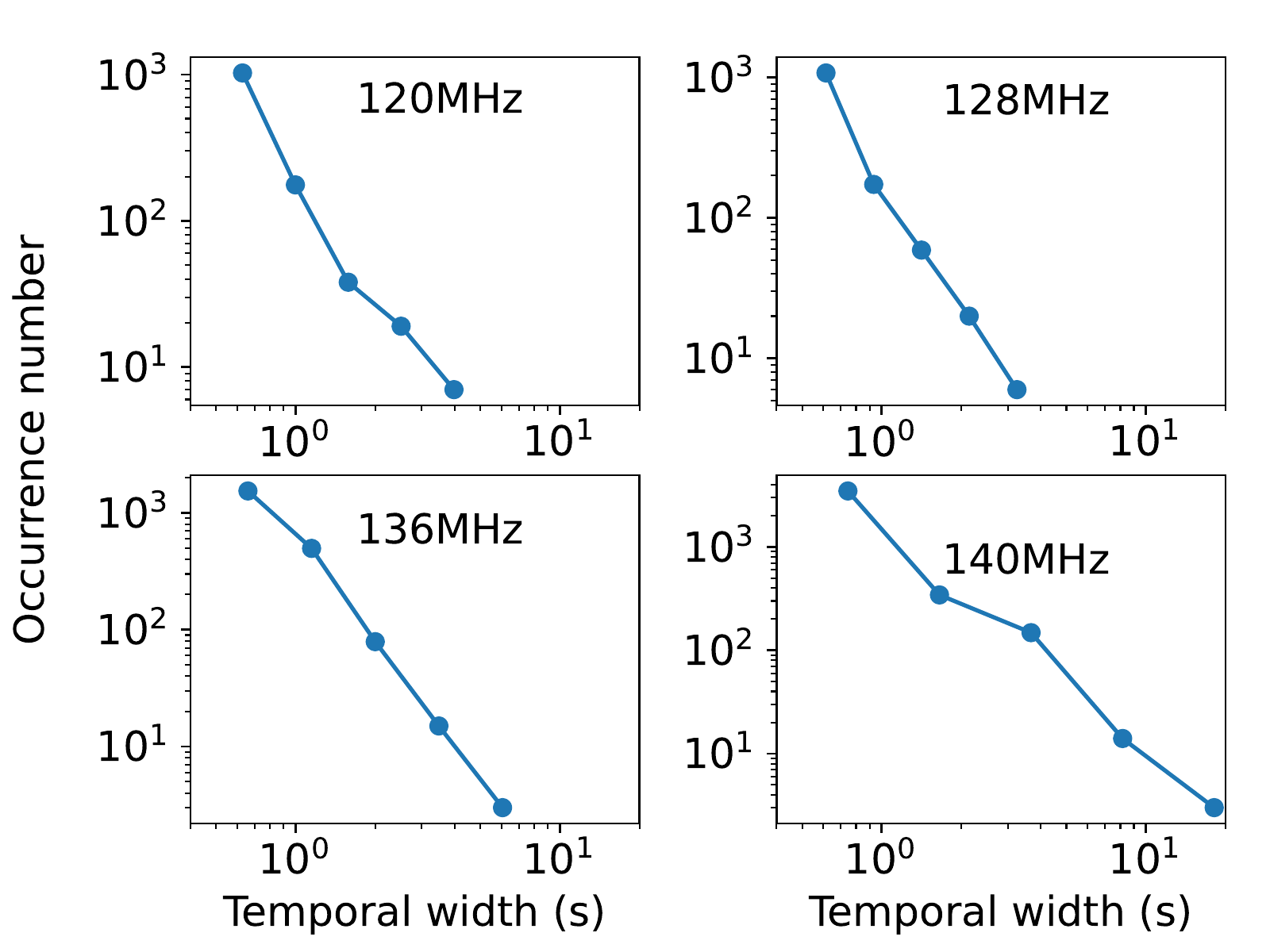}
    \caption{Temporal width distribution of the detected WINQSEs.}
    \label{fig:temporal_distribution}
\end{figure}

The temporal width of a WINQSE is defined as the time duration for which the value of a pixel in $I_{res}$ continuously exceeds the 3$\sigma$ threshold.
Figure \ref{fig:temporal_distribution} shows the temporal width distribution of the detected WINQSEs. The powerlaw index of the distributions is close to -2, very similar to that found in M20.
The bursts are highly impulsive with a steep powerlaw like distribution. 
The peak of the distribution lies at the instrumental resolution of 0.5 s, implying that most of the WINQSEs are unresolved in time.

\subsection{Wait time distribution}
\begin{figure}
    \centering
    \includegraphics[scale=0.5]{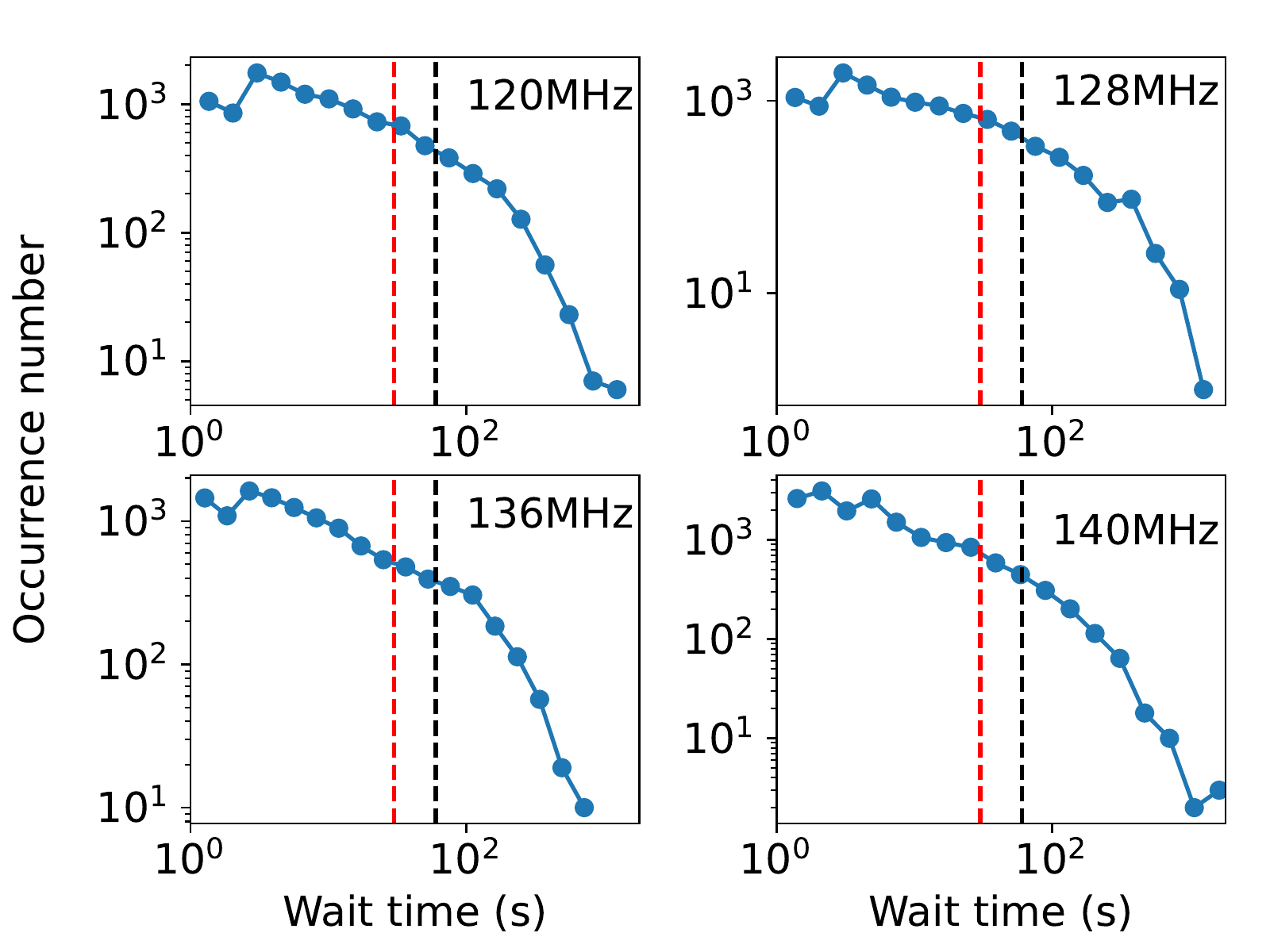}
    \caption{Wait time distribution of the detected WINQSEs. The red and black lines show the wait time till which the uncertainties due to incomplete sampling is less than 10\% and 20\% respectively.}
    \label{fig:wait_time_distribution}
\end{figure}

The wait time between WINQSEs is defined as the time interval between the successive times when the pixel value at a given pixel in $I_{res}$ exceeds the 3$\sigma$  threshold. The wait time distribution of detected WINQSEs is shown in Fig. \ref{fig:wait_time_distribution}. The wait time distribution is very similar to that found in M20. During the interpretation of the wait time distribution, a key point was however missed by M20. \citet{kivela2015} pointed out that the sharp cutoff at high wait times, like what is observed in Fig. \ref{fig:wait_time_distribution}, can arise because of data gaps and hence is not reliable. To take this into account, in Fig. \ref{fig:wait_time_distribution}, we draw red and black dashed lines to indicate the 10\% and 20\% error probabilities respectively. 
This implies that the data is sufficient to say that at low wait times the distribution follows a powerlaw, but the nature of the distribution at higher wait times cannot be constrained with these data. 
The powerlaw nature of wait time distribution is observed for solar flares and has been explained using a non-stationary Poisson model for solar flares \citep{aschwanden2010}.
It is possible that similar behaviour is also present for WINQSEs. M20 mention this aspect and alluded to the possibility that the observed wait time distribution might be severely affected by a lack of data. 
Based on the analysis here we conclude that this is indeed the case for the high wait time regime.

\subsection{Spatial distribution}

\begin{figure}
    \centering
    \includegraphics[trim={1cm 0 0 0},clip,scale=0.35]{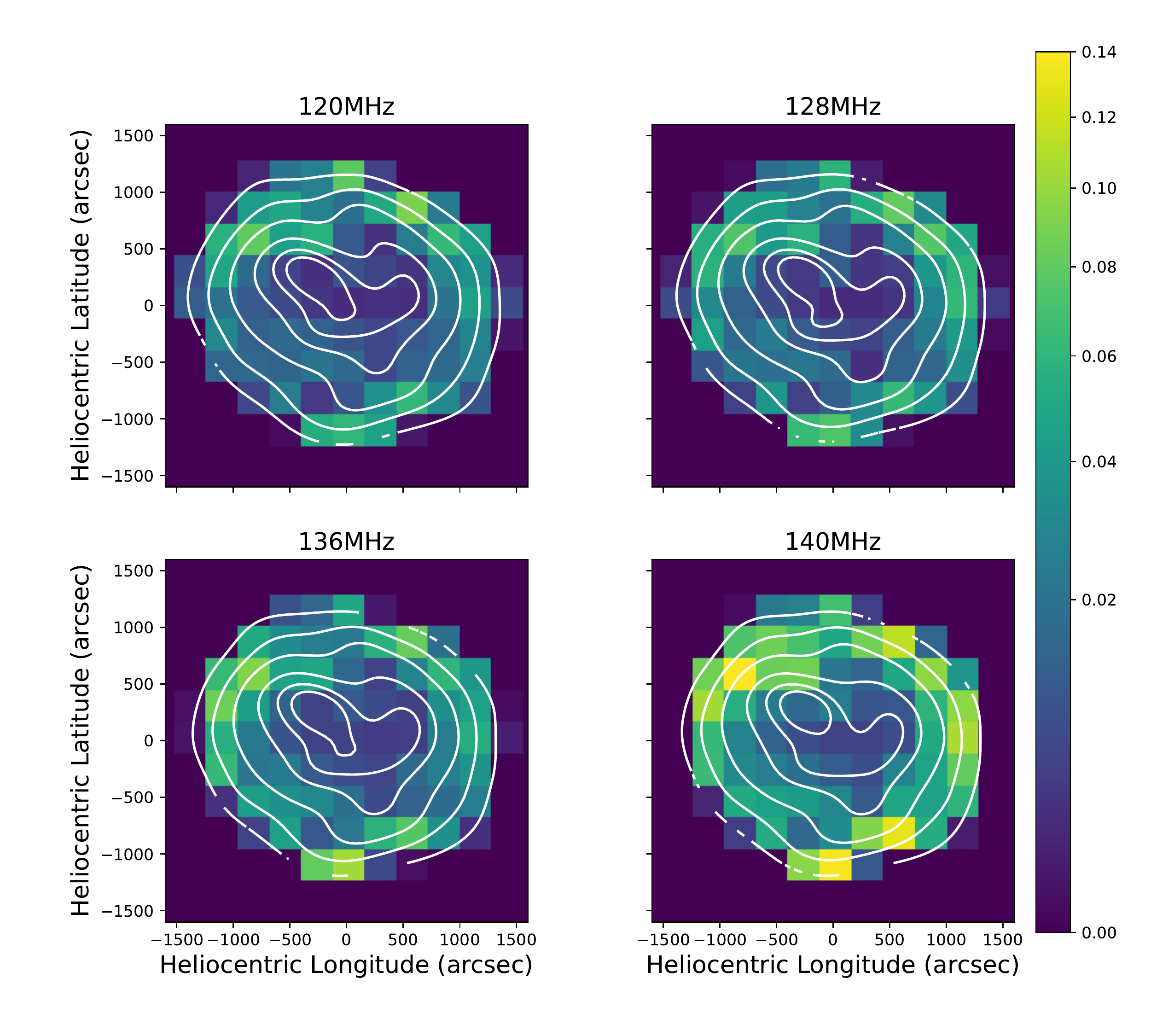}
    \caption{Fractional occupancy distribution of the detected WINQSEs. Contours of the median map at respective frequencies has been overlaid. The contour levels correspond to 0.2, 0.4, 0.6, 0.8, 0.9, 0.95 times the peak in the median map.
    }
    \label{fig:spatial_dist}
\end{figure}

A key result of M20 was that the WINQSEs are ubiquitous in nature. 
Figure \ref{fig:spatial_dist} shows the fractional occupancy of the detected WINQSEs over the radio Sun. 
The fractional occupancy is defined as the fraction of observation time for which WINQSEs were detected in a resolution element. 
There is no patch on the Sun where WINQSEs are absent.

\section{Discussion}

\subsection{Improving the data analysis technique over M20}

We have taken this opportunity to improve the analysis methodology to account for two low-level effects which were realized during the course of this work and are listed below:

\begin{enumerate}
    \item In M20 each five minute datachunk was processed independently, this can lead to small differences in absolute fluxscale between data chunks. At a few percent level, these differences are smaller than the uncertainty in the absolute flux density and usually too small to be of any consequence. In the present context, however, it is important to correct for them as they are comparable in strength to flux densities of WINQSEs. 
    As described in Sec. \ref{sec:analysis}, an additional step of forcing the median disc integrated solar flux density to a constant value has been introduced in the calibration process to ensure a constant fluxscale throughout the observing duration.
 
    \item During the calibration process, a varying number of antennas can get flagged for different time slices even at the same frequency. This can lead to small differences between the psfs for different images. To ensure that this does not affect the flux density estimated, all of the images were smoothed to the same coarser resolution for this work, as described in Sec. \ref{sec:analysis}.
\end{enumerate}

\subsection{Comparison with M20}
\label{subsec:M20comarison}

The results from this improved methodology confirm the key properties of WINQSEs reported by M20 -- weak flux densities, impulsive nature and ubiquitous presence on the quiet Sun during this period of very low solar activity. However, there are some significant quantitative differences which need some discussion.

\subsubsection{$\Delta F/F$ distribution of WINQSEs}

This work highlights that the distribution of $\Delta F/F$, can change significantly between observations.
At the low $\Delta F/F$ end, the distributions found by M20 tend to flatten out rather than drop, as seen in Fig. \ref{fig:deltaF_histogram}.
At the high $\Delta F/F$ end they show clear power law tails spanning more than an order of magnitude which are clearly missing in this work.

\begin{figure}
    \centering
    \includegraphics[trim={0 1cm 0 0},clip,scale=0.5]{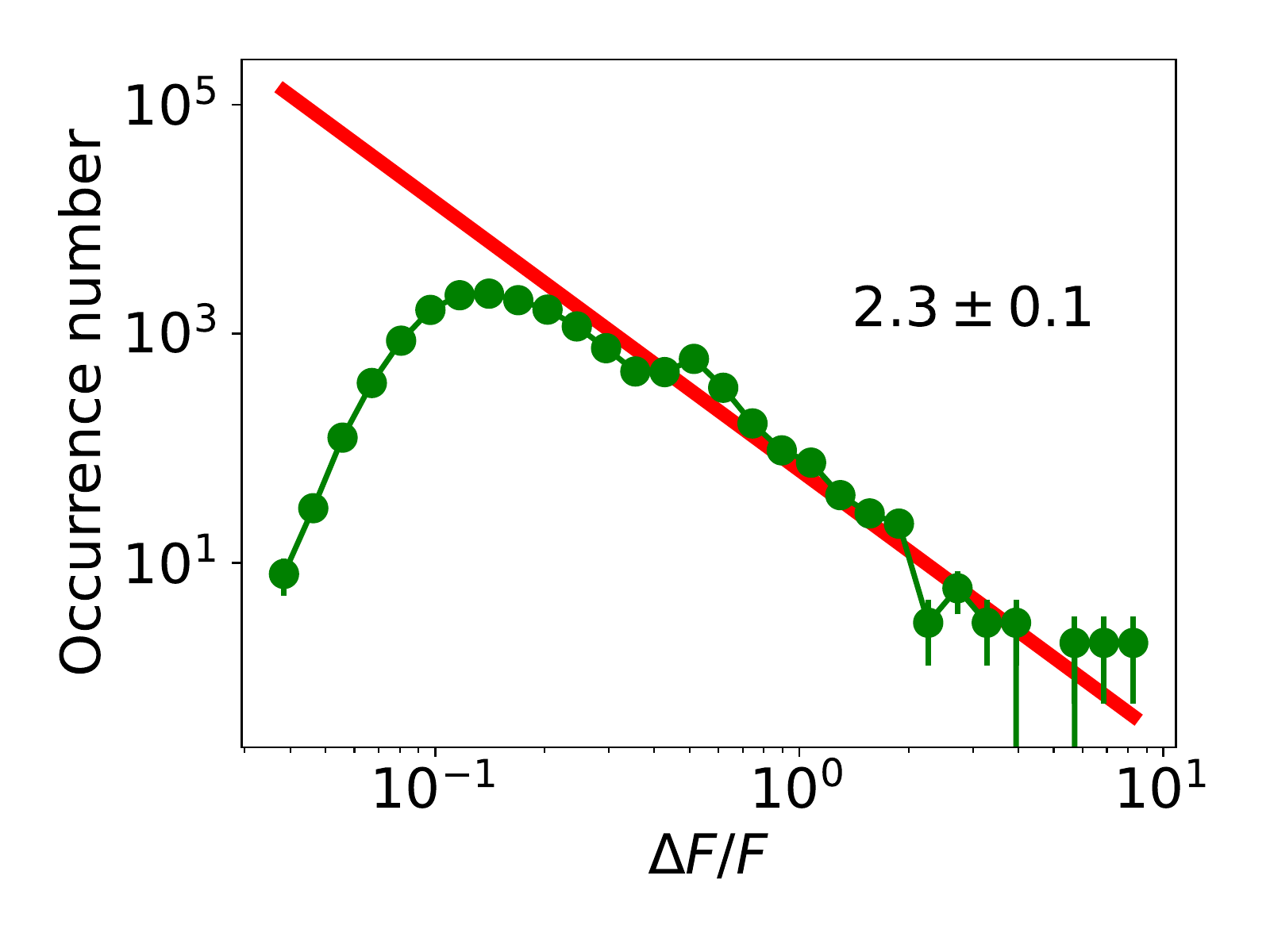}
    \caption{The data used here correspond to the 132 MHz data presented in M20.
    The $\Delta F/F$ histogram using data from M20 following the same procedure as that followed here. The red line shows the powerlaw fit to the data beyond the peak. The powerlaw index is mentioned in the figure.}
    \label{fig:compare_M20}
\end{figure}

For a given frequency, M20 plotted all $\Delta F/F$ values for all times and all the psf sized regions when presenting these histograms, irrespective of the signal-to-noise of the detection. 
Hence the low end of the $\Delta F/F$ distribution was dominated primarily by noise fluctuations. 
In contrast, we use an image noise based threshold on $\Delta F$ to limit ourselves only to reliable detections of WINQSEs.
This filters out the low signal-to-noise detections and explains the discrepancy between these works at the low end of the $\Delta F/F$ histograms. To substantiate this, we plot in Fig. \ref{fig:compare_M20} the data from an example frequency from M20 with the same thresholding technique as used here. 
It is evident that the low $\Delta F/F$ end of the histogram is now similar to the ones shown in Fig. \ref{fig:deltaF_histogram}, though it retains the powerlaw behaviour shown in M20 at the large $\Delta F/F$ end. The powerlaw index estimated is consistent with that reported in M20. Similar results were obtained at other frequencies as well.

\begin{figure*}
    \centering
    \includegraphics[scale=0.33]{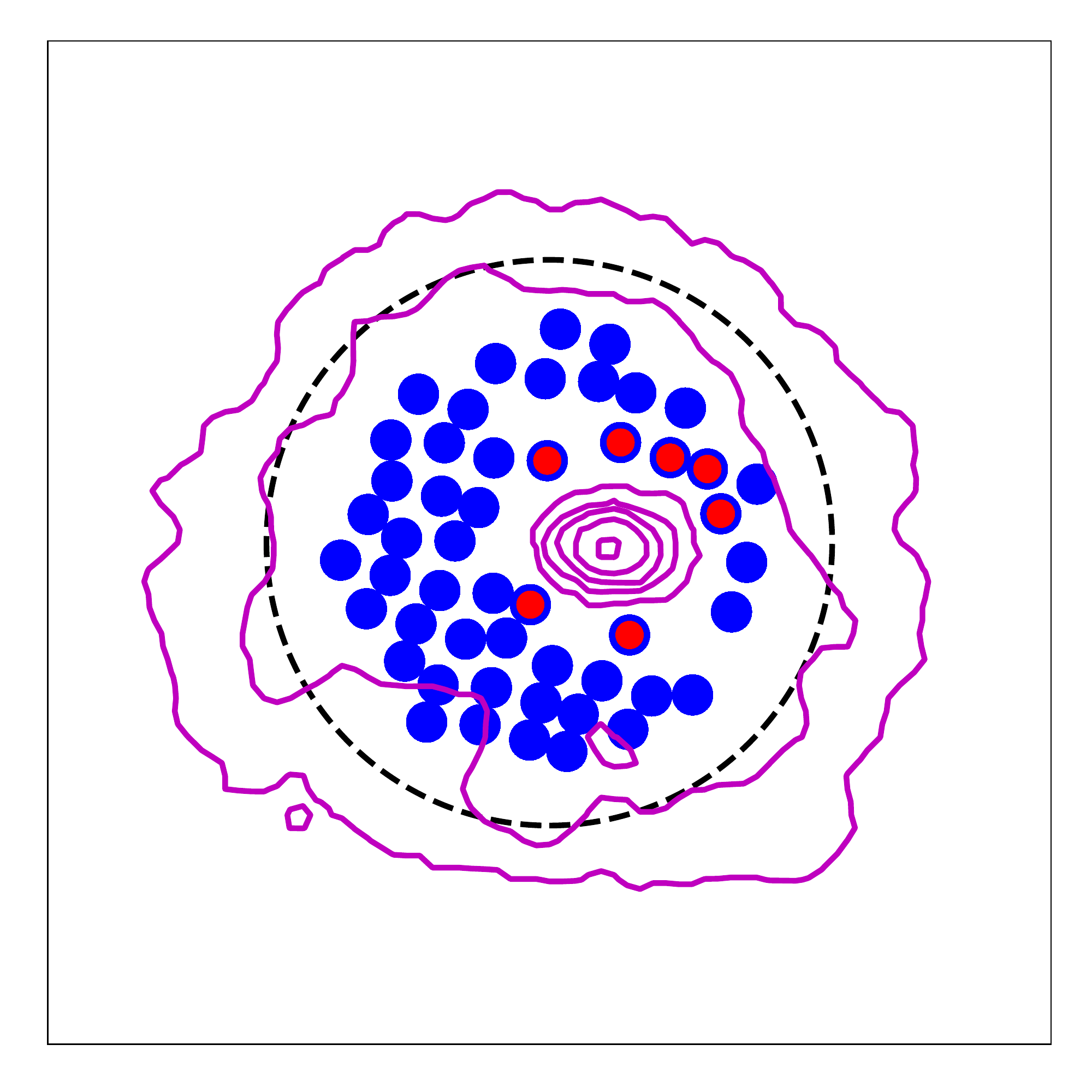}
    \includegraphics[trim={0.3cm 1cm 4.2cm 0},clip, scale=0.66]{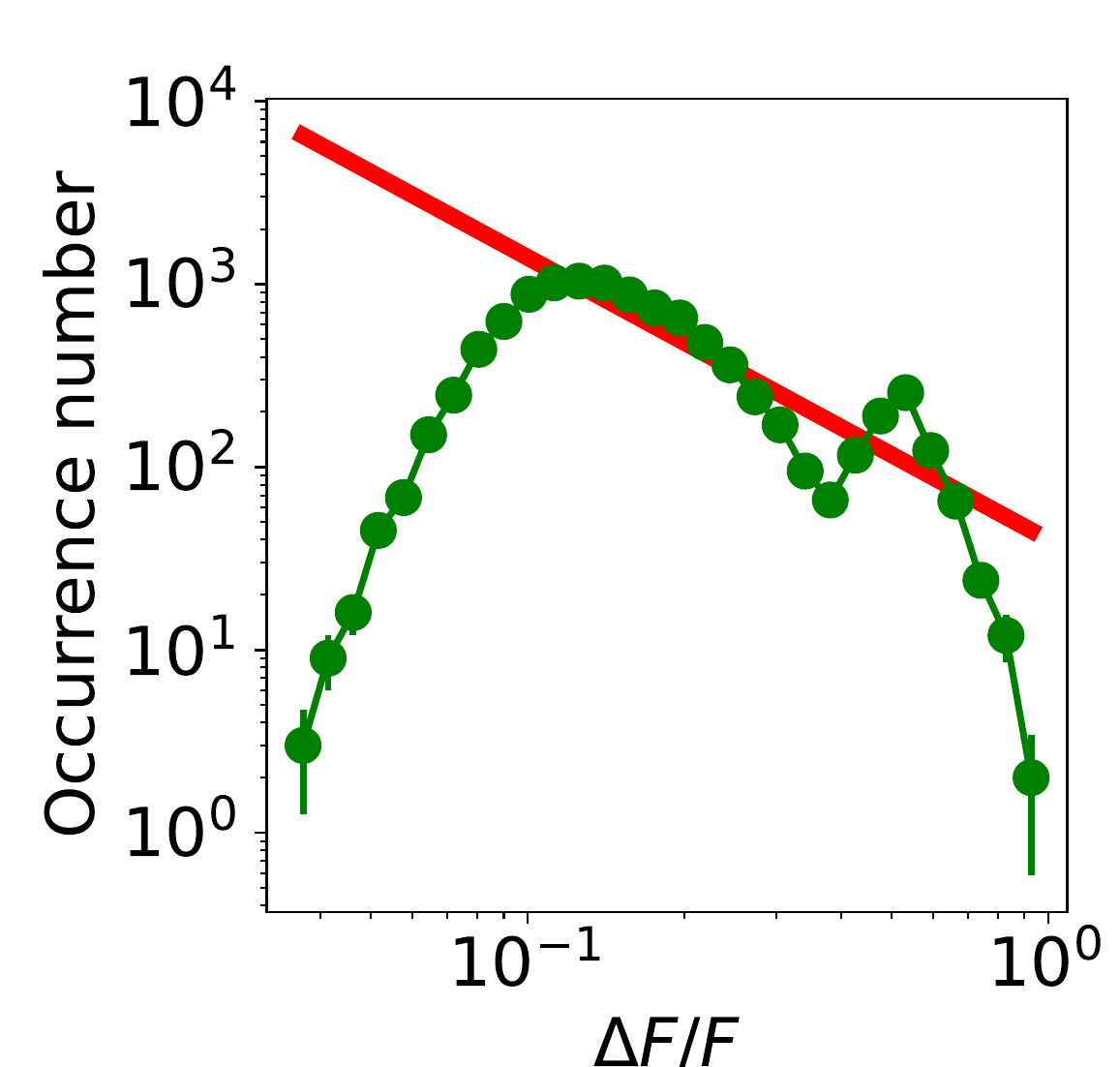}
    \caption{Left panel: The blue filled circles show the locations of all the psf sized regions used by M20 with WINQSEs detections. The regions which were determined by M20 to be affected by active region emission and excluded from their analysis have been excluded here as well. The red filled circles show the locations where the $\Delta F/F$ of the detected WINQSEs exceeds 1. The contours correspond to 0.01, 0.02, 0.04, 0.08, 0.16, 0.32, 0.64 times the peak of an example image at 132 MHz.  Histogram made using the same data as presented in Fig. \ref{fig:compare_M20} after removing contribution of regions which show at least one $\Delta F/F>1$.}
    \label{fig:m20_exclude_active_regions_close}
\end{figure*}

Though M20 had attempted to avoid contamination from the only active region present on the Sun during their observations, we find that despite their efforts the high $\Delta F/F$ tail  observed in the histograms obtained in M20 arises due to this lone active region.
To substantiate this, the left panel of Fig. \ref{fig:m20_exclude_active_regions_close} shows the location of all WINQSEs detected in M20 at 132 MHz, overlaid on an example radio image at the same frequency.
The regions are colored blue (red) if they host at least one WINQSE with $\Delta F/F>0$ ($>1$). 
It is evident that while WINQSEs are distributed all over the Sun, those with $\Delta F/F>1$ are clustered around the active region and not a single one of them is present far away from it.

Active regions have very complex magnetic fields and are sites of continuous magnetic reconnections \citep[e.g.][]{mondal2021a}. The complexity of the magnetic field is expected to decrease smoothly as the distance from the active region increases. Consequently, the strength and the number of magnetic reconnections taking place would also decrease in a similar smooth manner. We hypothesise this to be the reason for regions showing the presence of strong WINQSEs in M20 to lie close to the active region. Additionally, such a region is expected not only to produce strong WINQSEs, but also to give rise to a larger number of weaker WINQSEs, as compared to regions far away from the active region. 
Hence we hypothesise this to be
the reason behind the presence (absence) of powerlaw behaviour in the $\Delta F/F$ distribution of the detected WINQSEs in M20 (this work). To corroborate this hypothesis, the right panel of Fig. \ref{fig:m20_exclude_active_regions_close} presents the $\Delta F/F$ distribution after removing the contribution of the regions shown in red in the left panel. The red line shows the straight line fitted to the same range of $\Delta F/F$ used for the fit in Fig. \ref{fig:compare_M20}. It is evident that the powerlaw is no longer a good representation of the distribution. It is also observed that there is hump-like feature around $\Delta F/F \approx 0.5$. 
The signature of this hump is also observed at a similar location in Fig. \ref{fig:compare_M20} in this work and Fig 4 of M20. 
Understanding the exact details of this feature is beyond the scope of the present work.
This observation of the transition from a powerlaw behaviour towards a more lognormal-like nature suggests that as the solar activity increases, more active regions are observed, leading to a larger number of stronger WINQSEs. 
Hence with increasing solar activity a transition from a lognormal nature to a powerlaw behaviour in the $\Delta F/F$ distribution can be expected. This prediction should be tested in future with data spanning a large part of a solar cycle.

Note that the peaks in the $\Delta F/F$ histograms lie at lower values in the current data as compared to M20. We attribute this to a combination of exceptionally quiet solar conditions and lower image noise characteristics delivered by the improved calibration employed here.

\subsubsection{Fractional occupancy of WINQSEs}

The median fractional occupancies obtained from the data presented here are 0.017, 0.021, 0.024 and 0.045 at 120, 128, 136 and 146 MHz respectively. 
In comparison, at similar frequencies of 120 and 132 MHz, M20 had found higher median fractional occupancies of 0.07 and 0.06 respectively.
There are at least four known factors which contribute to a higher fractional occupancy for M20.
The first is the improved analysis methodology used here which rigorously rejects low signal-to-noise detections. This was not the case with M20 and hence some spurious noise peaks can get counted as WINQSEs inflating the true fractional occupancy number. Secondly, M20 computed a single median for the entire duration, in contrast to the piece-wise median estimated here. The appendix shows the results obtained using a single median and it is evident that the results thus obtained are closer to that obtained by M20 (3-5\%).
The third important reason is the more active state of the Sun during M20, some evidence and arguments for contributions from which have already been presented.

{
\subsection{Comparison with S22}
\label{sec:comp_S22}

As mentioned earlier, S22 have recently reported the detection of WINQSEs using a very different technique, namely the {\em visibility subtraction} technique. For completeness, we briefly describe the technique first and then present a comparison of their technique and the results with the present work. 

Similar to the technique used here, S22 also assumes that there is a {\em slowly varying} background solar radio emission on which are the superposed the impulsive WINQSEs. 
While we separate the slowly varying and impulsive parts of emission in the image domain by estimating and subtracting a median image, S22 does so in the visibility domain.
For each baseline, they compute the median complex visibility over a 15 second running window and assume that the median represents the visibility from the slowly varying Sun for that baseline for the instant centered on that time interval. 
In principle, subtraction in the visibility domain and subtraction in the image domain should be identical due to the Fourier relationship between them.
However, in practice, there can be differences due to non-linear nature of deconvolution which is used to produce the radio images. 
The subtraction in visibility domain is expected to perform better if there are significant deconvolution errors in the images. 
A disadvantage of the visibility subtraction technique, however, is that the Fourier component sampled by a given baseline changes as a function of time. For a source with the complex morphology like the Sun, the emission morphology can itself give rise to variations in the observed visbilities on a given baseline even if the source were time invariant.
To limit the magnitudes of these intrinsic changes in measured visibilities, S22 used a rather short duration of 15 s over which to compute the median, while our technique was able to use the full 4 min duration of the scan to compute the median image. 
An implication is that S22 would miss the longer duration events. While this will bias any studies of the temporal width distribution done using this technique, given the very steep slope of this distribution, however, the number of events missed will be a very small fraction of the total number.

The solar images used in this work were of high quality and no deconvolution errors or artefacts were evident. In addition to the excellent uv-coverage, precise calibration performed using AIRCARS \citep{mondal2019} and the absence of any strong source on the solar disc, all have contributed to ensuring a high fidelity images. 
This has allowed us to use the image subtraction technique with a high confidence and enabled the detection of these WINQSEs.
}

\subsection{Need for a better WINQSE detection technique}
\label{subsec:shortcomings}

A shortcoming of both this work and M20 stems from the way WINQSEs are detected. We have searched for WINQSEs in discrete psf-sized regions and classified the transients as true detections if their flux density exceeds some threshold. If a WINQSE is larger than the psf, then it will contribute flux density to multiple regions and hence can lead to over-counting of the number of WINQSEs. 
On the other hand, since the psf is circular/elliptical in nature, the regions used here and in M20 do not form a closed packed structure. Hence some part of the Sun is not covered by these regions and could lead to some WINQSEs being missed. 
Additionally, when the spatial location of a WINQSE does not coincide with the centre of some psf sized region, the observed increase in flux density of the region due to the WINQSE is lower than its true flux density. This will lead to instances when the observed flux density will fall short of the detection threshold due to this artificial reduction. 
Due to these effects, it is non-trivial to estimate the true number of WINQSEs present in the data.
To address this shortcoming, we have recently developed a machine learning based technique to reliably identify and characterize WINQSEs in solar radio images. 
This work will be presented in an upcoming paper \citep{bawaji2022}. 

An additional shortcoming of both these works stems from the way the median is computed. M20 had computed a single median map for the entire time duration studied, whereas here we have computed a median map for every five minutes of data. Both of these methods have their own strengths and weaknesses. 
For example, while calculating an independent median for every five minutes of data, can reduce some systematic calibration artefacts, if present, it can also produce spurious WINQSEs at the five minute data boundaries if for that data chunk and pixel, the lightcurve has a smoothly decreasing or increasing trend. We have repeated the entire analysis using a single median map and, as expected, we find that the number of WINQSEs found in this analysis is significantly larger than when done using independent medians for each data chunk. We are not aware of any systematic effects in the data which can change the flux densities in different regions of the Sun while maintaining the integrated flux at all times.
Choosing to err on the side of caution, we have presented the results corresponding to the medians computed every five minutes in the preceding sections. The statistical properties of WINQSEs obtained from the single median analysis are very similar and are presented in the Appendix. While the results are consistent, we realize the need to improve WINQSEs detection strategy beyond what has been employed here. 
Work on this front is already underway.




\subsection{A dynamic middle corona}

Radio observations at the small range of frequencies used here probe coronal height of $\sim 1.4\ (2.0)\ R_\odot$, under the standard assumption that the emission arises due to plasma emission mechanism at the fundamental (harmonic) and using well established coronal electron density models.
The observations in M20, \citet{sharma2022} and this work imply that energetically weak reconnections are ever-present even in the quiet Sun regions at these coronal heights.
These reconnections accelerate electrons which then emit in the radio band. 
Due to instrumental limitations at high energies (EUV and X-rays), these coronal heights are typically accessible only at metrewave radio frequencies. 
The bulk of the radio studies in the past have focused on strong events, again primarily due to the instrumental limitations in radio bands. 
So there have been rather few investigations of weak or level emissions from the middle coronal heights.
As more capable instruments are becoming available, the evidence for middle coronal heights to also be a dynamic place is slowly building.
Recently, gradual bulk flows originating in the middle corona have been observed using the Solar Ultraviolet Imager (SUVI) onboard the Geostationary Operational Environmental Satellite \citep{seaton2021}.
The authors suggest that this can happen due to magnetic reconnection processes higher in the corona.
While a direct comparison between these EUV emissions and WINQSEs is beyond the scope of this work, these disparate observations paint a consistent picture of a highly dynamic middle solar corona even during extremely quiet times.


\section{Conclusion}

M20 was the first work to present ubiquitous detection of WINQSEs. Due to the significant potential implications of this work on coronal heating, and presence of nonthermal particles in the solar corona in general, it is very important to verify their presence under a variety of solar conditions, take a closer look at the analysis and characterise them better.   
With this motivation, this work forms the next step in this progression.
We have examined data from extremely quiet solar conditions, improved on the methodology used in M20 and presented robust detection of large numbers of WINQSEs. 
This work corroborates the key findings of M20 -- their ubiquity and impulsive nature.
While significant differences between the distribution of $\Delta F/F$ presented here and M20 are observed, we show that they can be justifiably attributed to differences in methodology and variations in the level of solar activity.
Parallel to this effort, independent techniques have been developed to identify and characterise WINQSEs \citep{bawaji2022,sharma2022} and they all paint a consistent picture.
Together these works place the presence of WINQSEs on a firm pedestal.  
We hope that this will provide the motivation to the wider community to understand them in greater detail. 

\facilities{Murchison Widefield Array (MWA) \citep{lonsdale2009,tingay2013}}

\software{astropy \citep{astropy:2013,astropy:2018}, matplotlib \citep{Hunter:2007}, Numpy \citep{harris2020array}, SciPy \citep{2020SciPy-NMeth}, CASA \citep{mcmullin2007}}

\begin{acknowledgments}

This scientific work makes use of the Murchison Radio-astronomy Observatory, operated by CSIRO. We acknowledge the Wajarri Yamatji people as the traditional owners of the Observatory site. Support for the operation of the MWA is provided by the Australian Government (NCRIS), under a contract to Curtin University administered by Astronomy Australia Limited. We acknowledge the Pawsey Supercomputing Centre which is supported by the Western Australian and Australian Governments. We thank the anonymous referee for their comments, which have improved this paper significantly.
SM acknowledges partial support by USA NSF grant AGS-1654382 to the New Jersey Institute of Technology.
AB and DO acknowledge support of the Department of Atomic Energy, Government of India, under the project no. 12-R\&D-TFR-5.02-0700. 

\end{acknowledgments}

\appendix

As discussed in Sec. \ref{subsec:shortcomings}, the number of WINQSEs detected depends on the choice of the median map used.
Results presented in the paper are based on the use of an independent median computed for each five minute dataset.
For completeness and comparison, here we present results (analogs of Figs. 3--6) obtained using a single median computed over the entire observing duration as was done by \citet{mondal2020}.
As expected the number of WINQSEs detected here are larger than using a median computed every five minutes.

\begin{figure}
    \centering
    \includegraphics[scale=0.5]{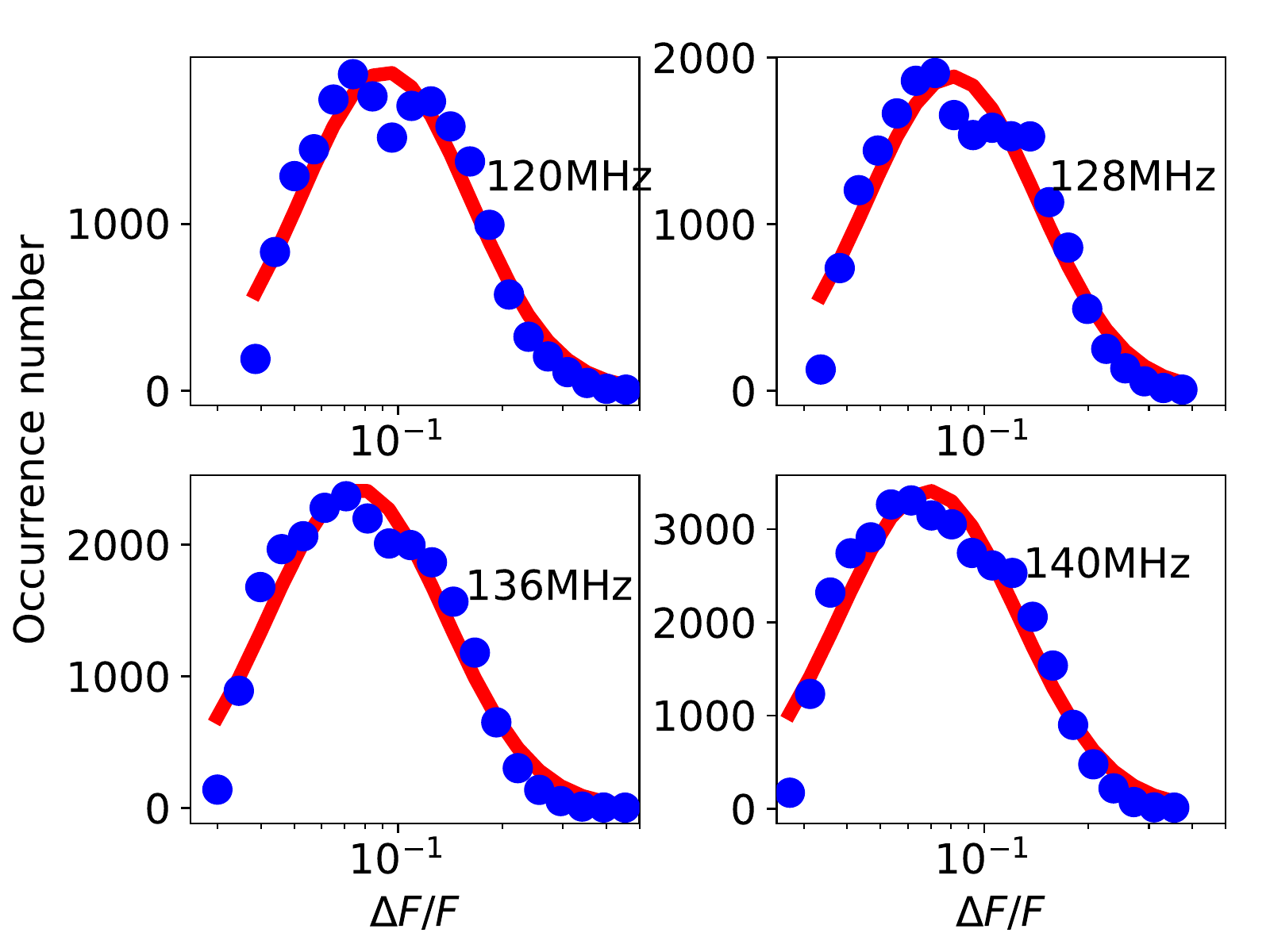}
    \caption{Histogram of $\Delta F/F$ using a median computed over the entire duration of the data is shown using the blue circles. The red line shows the lognormal fit to these data. The observation frequencies are written in each panel.}
    \label{fig:deltaF_histogram1}
\end{figure}

\begin{figure}
    \centering
    \includegraphics[scale=0.5]{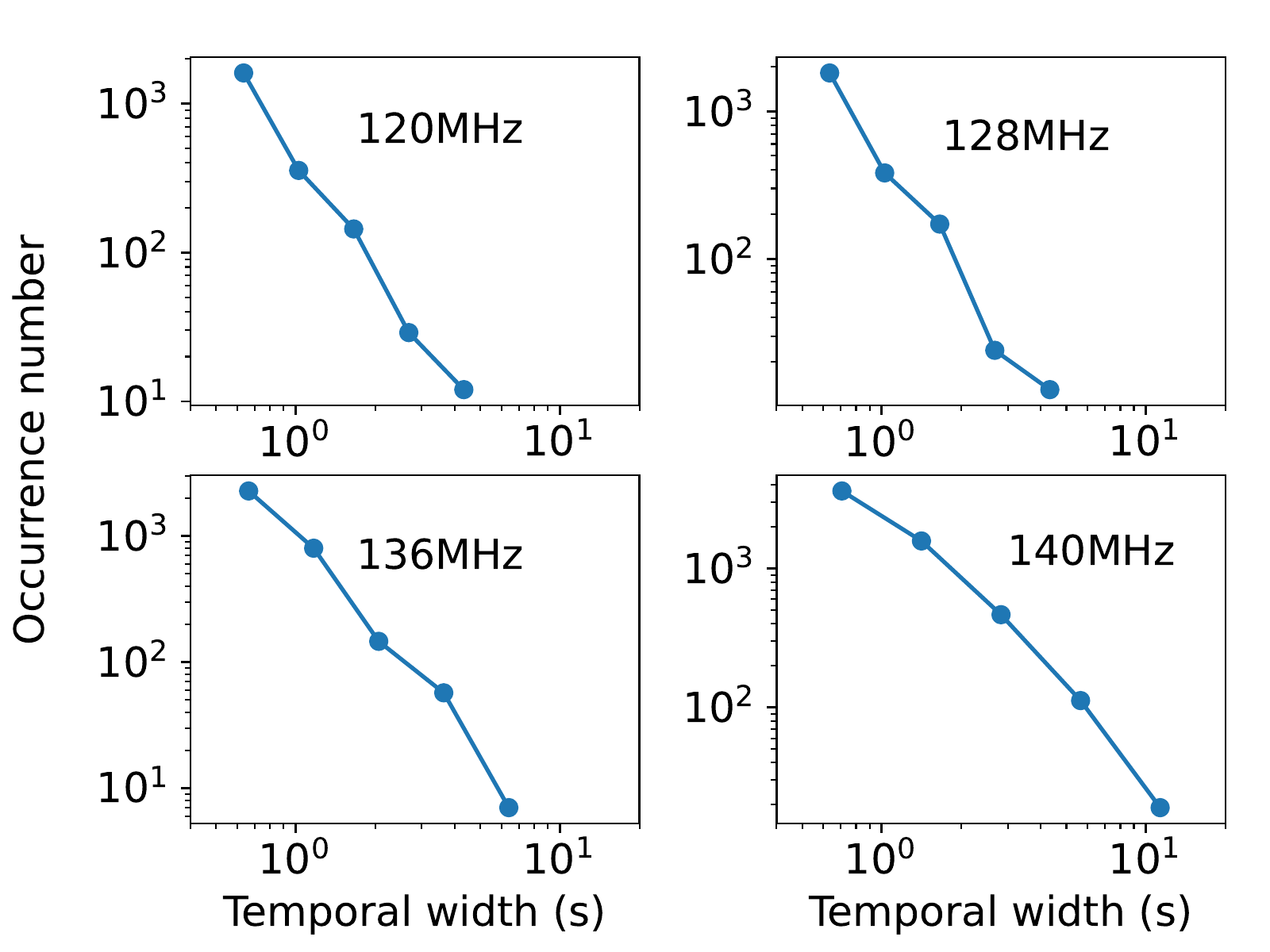}
    \caption{Temporal width distribution of the detected WINQSEs using a median computed over the entire duration of the data.}
    \label{fig:temporal_distribution1}
\end{figure}

\begin{figure}
    \centering
    \includegraphics[scale=0.5]{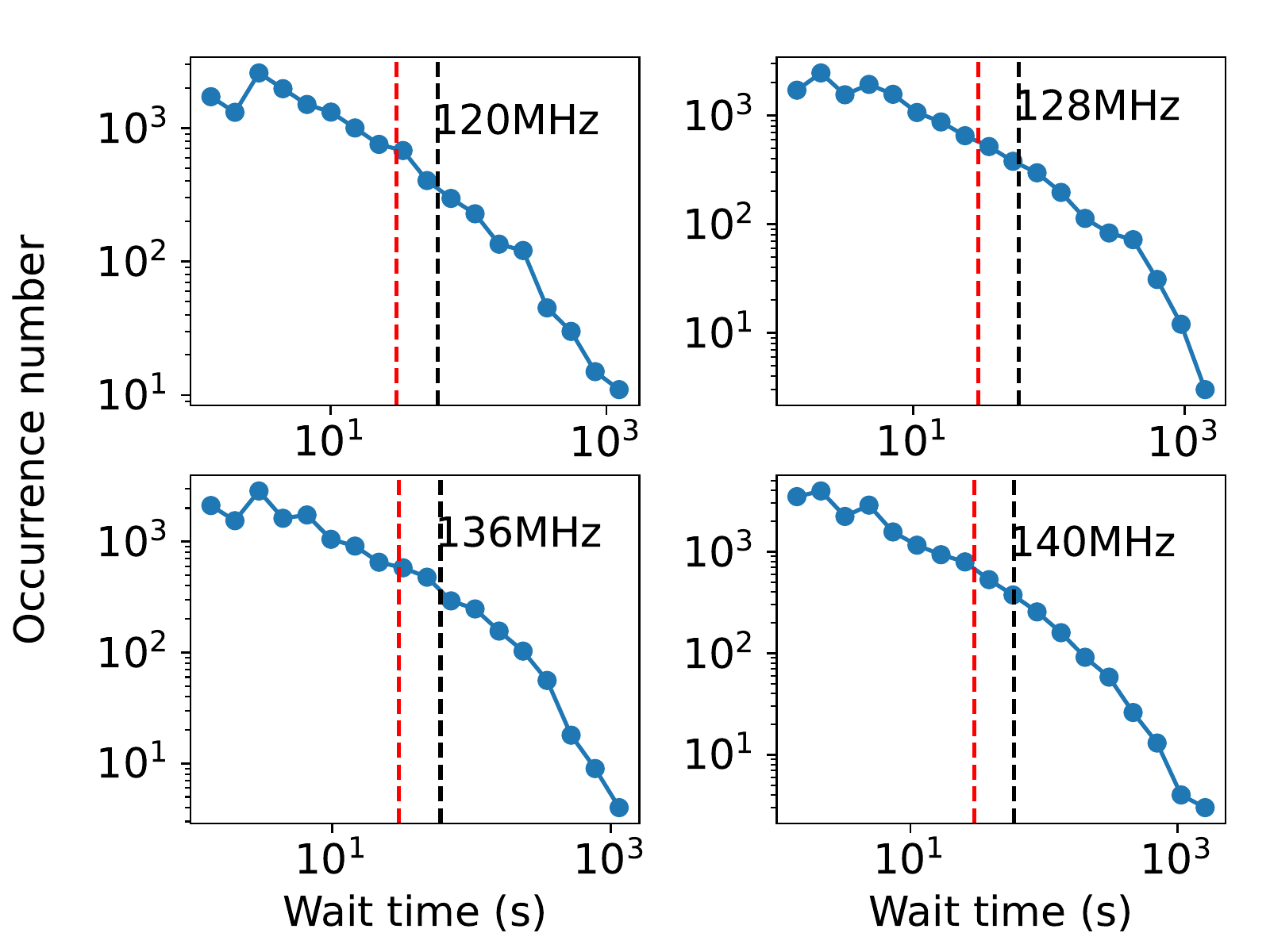}
    \caption{Wait time distribution of the detected WINQSEs using a median computed over the entire duration of the data. The red and black lines show the wait time till which the uncertainties due to incomplete sampling is less than 10\% and 20\% respectively.}
    \label{fig:wait_time_distribution1}
\end{figure}

\begin{figure}
    \centering
    \includegraphics[trim={1cm 0 0 0},clip,scale=0.45]{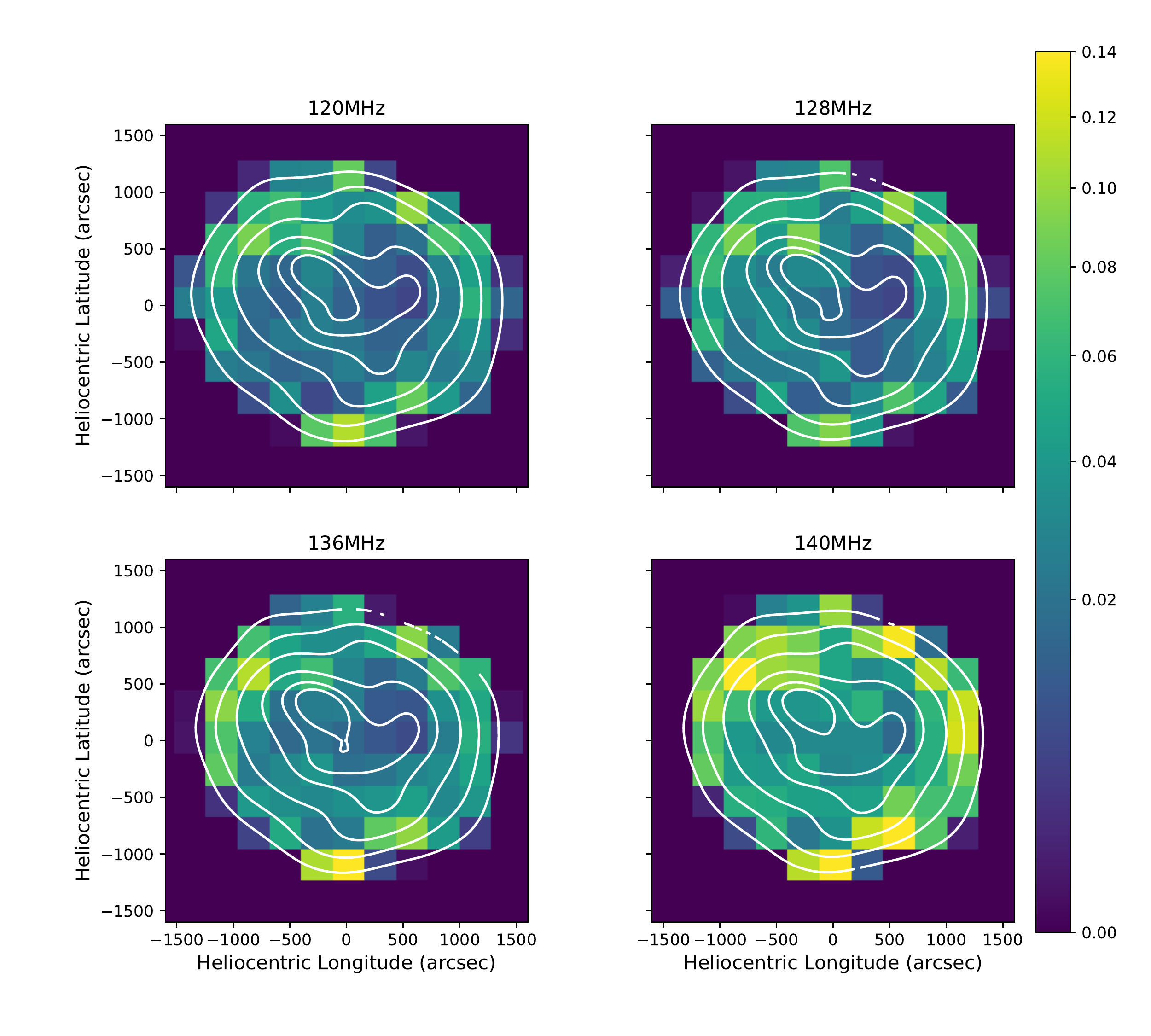}
    \caption{Fractional occupancy distribution of the detected WINQSEs using a median computed over the entire duration of the data.}
    \label{fig:spatial_dist1}
\end{figure}

\bibliography{sample63}{}

\begin{thebibliography}{}
\expandafter\ifx\csname natexlab\endcsname\relax\def\natexlab#1{#1}\fi
\providecommand{\url}[1]{\href{#1}{#1}}
\providecommand{\dodoi}[1]{doi:~\href{http://doi.org/#1}{\nolinkurl{#1}}}
\providecommand{\doeprint}[1]{\href{http://ascl.net/#1}{\nolinkurl{http://ascl.net/#1}}}
\providecommand{\doarXiv}[1]{\href{https://arxiv.org/abs/#1}{\nolinkurl{https://arxiv.org/abs/#1}}}

\bibitem[{{Aschwanden} \& {McTiernan}(2010)}]{aschwanden2010}
{Aschwanden}, M.~J., \& {McTiernan}, J.~M. 2010, \apj, 717, 683,
  \dodoi{10.1088/0004-637X/717/2/683}

\bibitem[{{Astropy Collaboration} {et~al.}(2013){Astropy Collaboration},
  {Robitaille}, {Tollerud}, {Greenfield}, {Droettboom}, {Bray}, {Aldcroft},
  {Davis}, {Ginsburg}, {Price-Whelan}, {Kerzendorf}, {Conley}, {Crighton},
  {Barbary}, {Muna}, {Ferguson}, {Grollier}, {Parikh}, {Nair}, {Unther},
  {Deil}, {Woillez}, {Conseil}, {Kramer}, {Turner}, {Singer}, {Fox}, {Weaver},
  {Zabalza}, {Edwards}, {Azalee Bostroem}, {Burke}, {Casey}, {Crawford},
  {Dencheva}, {Ely}, {Jenness}, {Labrie}, {Lim}, {Pierfederici}, {Pontzen},
  {Ptak}, {Refsdal}, {Servillat}, \& {Streicher}}]{astropy:2013}
{Astropy Collaboration}, {Robitaille}, T.~P., {Tollerud}, E.~J., {et~al.} 2013,
  \aap, 558, A33, \dodoi{10.1051/0004-6361/201322068}

\bibitem[{{Astropy Collaboration} {et~al.}(2018){Astropy Collaboration},
  {Price-Whelan}, {Sip{\H{o}}cz}, {G{\"u}nther}, {Lim}, {Crawford}, {Conseil},
  {Shupe}, {Craig}, {Dencheva}, {Ginsburg}, {Vand erPlas}, {Bradley},
  {P{\'e}rez-Su{\'a}rez}, {de Val-Borro}, {Aldcroft}, {Cruz}, {Robitaille},
  {Tollerud}, {Ardelean}, {Babej}, {Bach}, {Bachetti}, {Bakanov}, {Bamford},
  {Barentsen}, {Barmby}, {Baumbach}, {Berry}, {Biscani}, {Boquien}, {Bostroem},
  {Bouma}, {Brammer}, {Bray}, {Breytenbach}, {Buddelmeijer}, {Burke},
  {Calderone}, {Cano Rodr{\'\i}guez}, {Cara}, {Cardoso}, {Cheedella}, {Copin},
  {Corrales}, {Crichton}, {D'Avella}, {Deil}, {Depagne}, {Dietrich}, {Donath},
  {Droettboom}, {Earl}, {Erben}, {Fabbro}, {Ferreira}, {Finethy}, {Fox},
  {Garrison}, {Gibbons}, {Goldstein}, {Gommers}, {Greco}, {Greenfield},
  {Groener}, {Grollier}, {Hagen}, {Hirst}, {Homeier}, {Horton}, {Hosseinzadeh},
  {Hu}, {Hunkeler}, {Ivezi{\'c}}, {Jain}, {Jenness}, {Kanarek}, {Kendrew},
  {Kern}, {Kerzendorf}, {Khvalko}, {King}, {Kirkby}, {Kulkarni}, {Kumar},
  {Lee}, {Lenz}, {Littlefair}, {Ma}, {Macleod}, {Mastropietro}, {McCully},
  {Montagnac}, {Morris}, {Mueller}, {Mumford}, {Muna}, {Murphy}, {Nelson},
  {Nguyen}, {Ninan}, {N{\"o}the}, {Ogaz}, {Oh}, {Parejko}, {Parley}, {Pascual},
  {Patil}, {Patil}, {Plunkett}, {Prochaska}, {Rastogi}, {Reddy Janga},
  {Sabater}, {Sakurikar}, {Seifert}, {Sherbert}, {Sherwood-Taylor}, {Shih},
  {Sick}, {Silbiger}, {Singanamalla}, {Singer}, {Sladen}, {Sooley},
  {Sornarajah}, {Streicher}, {Teuben}, {Thomas}, {Tremblay}, {Turner},
  {Terr{\'o}n}, {van Kerkwijk}, {de la Vega}, {Watkins}, {Weaver}, {Whitmore},
  {Woillez}, {Zabalza}, \& {Astropy Contributors}}]{astropy:2018}
{Astropy Collaboration}, {Price-Whelan}, A.~M., {Sip{\H{o}}cz}, B.~M., {et~al.}
  2018, \aj, 156, 123, \dodoi{10.3847/1538-3881/aabc4f}

\bibitem[{{Bain} \& {Fletcher}(2009)}]{bain2009}
{Bain}, H.~M., \& {Fletcher}, L. 2009, \aap, 508, 1443,
  \dodoi{10.1051/0004-6361/200911876}

\bibitem[{{Bawaji} {et~al.}(2022){Bawaji}, {Alam}, {Mondal}, \&
  {Oberoi}}]{bawaji2022}
{Bawaji}, S., {Alam}, U., {Mondal}, S., \& {Oberoi}, D. 2022, in Astronomical
  Society of the Pacific Conference Series, Vol. 532, Astronomical Society of
  the Pacific Conference Series, ed. J.~E. {Ruiz}, F.~{Pierfedereci}, \&
  P.~{Teuben}, 211.
\newblock \doarXiv{2103.05371}

\bibitem[{{Che}(2018)}]{che2018}
{Che}, H. 2018, in Journal of Physics Conference Series, Vol. 1100, Journal of
  Physics Conference Series, 012005, \dodoi{10.1088/1742-6596/1100/1/012005}

\bibitem[{Harris {et~al.}(2020)Harris, Millman, van~der Walt, Gommers,
  Virtanen, Cournapeau, Wieser, Taylor, Berg, Smith, Kern, Picus, Hoyer, van
  Kerkwijk, Brett, Haldane, del R{\'{i}}o, Wiebe, Peterson,
  G{\'{e}}rard-Marchant, Sheppard, Reddy, Weckesser, Abbasi, Gohlke, \&
  Oliphant}]{harris2020array}
Harris, C.~R., Millman, K.~J., van~der Walt, S.~J., {et~al.} 2020, Nature, 585,
  357, \dodoi{10.1038/s41586-020-2649-2}

\bibitem[{Hunter(2007)}]{Hunter:2007}
Hunter, J.~D. 2007, Computing in Science \& Engineering, 9, 90,
  \dodoi{10.1109/MCSE.2007.55}

\bibitem[{{Kansabanik} {et~al.}(2022{\natexlab{a}}){Kansabanik}, {Mondal},
  {Oberoi}, {Biswas}, \& {Bhunia}}]{kansabanik2022b}
{Kansabanik}, D., {Mondal}, S., {Oberoi}, D., {Biswas}, A., \& {Bhunia}, S.
  2022{\natexlab{a}}, \apj, 927, 17, \dodoi{10.3847/1538-4357/ac4bba}

\bibitem[{{Kansabanik} {et~al.}(2022{\natexlab{b}}){Kansabanik}, {Oberoi}, \&
  {Mondal}}]{kansabanik2022a}
{Kansabanik}, D., {Oberoi}, D., \& {Mondal}, S. 2022{\natexlab{b}}, \apj, 932,
  110, \dodoi{10.3847/1538-4357/ac6758}

\bibitem[{Kivel\"a \& Porter(2015)}]{kivela2015}
Kivel\"a, M., \& Porter, M.~A. 2015, Phys. Rev. E, 92, 052813,
  \dodoi{10.1103/PhysRevE.92.052813}

\bibitem[{{Kontar} {et~al.}(2017){Kontar}, {Yu}, {Kuznetsov}, {Emslie},
  {Alcock}, {Jeffrey}, {Melnik}, {Bian}, \& {Subramanian}}]{kontar2017}
{Kontar}, E.~P., {Yu}, S., {Kuznetsov}, A.~A., {et~al.} 2017, Nature
  Communications, 8, 1515, \dodoi{10.1038/s41467-017-01307-8}

\bibitem[{{Krucker} {et~al.}(2011){Krucker}, {Kontar}, {Christe}, {Glesener},
  \& {Lin}}]{krucker2011}
{Krucker}, S., {Kontar}, E.~P., {Christe}, S., {Glesener}, L., \& {Lin}, R.~P.
  2011, \apj, 742, 82, \dodoi{10.1088/0004-637X/742/2/82}

\bibitem[{Lonsdale {et~al.}(2009)Lonsdale, Cappallo, Morales, Briggs,
  Benkevitch, Bowman, Bunton, Burns, Corey, DeSouza, Doeleman, Derome,
  Deshpande, Gopala, Greenhill, Herne, Hewitt, Kamini, Kasper, Kincaid, Kocz,
  Kowald, Kratzenberg, Kumar, Lynch, Madhavi, Matejek, Mitchell, Morgan,
  Oberoi, Ord, Pathikulangara, Prabu, Rogers, Roshi, Salah, Sault, Shankar,
  Srivani, Stevens, Tingay, Vaccarella, Waterson, Wayth, Webster, Whitney,
  Williams, \& Williams}]{lonsdale2009}
Lonsdale, C.~J., Cappallo, R.~J., Morales, M.~F., {et~al.} 2009, Proceedings of
  the IEEE, 97, 1497, \dodoi{10.1109/JPROC.2009.2017564}

\bibitem[{{McLean} \& {Labrum}(1985)}]{mclean1985}
{McLean}, D.~J., \& {Labrum}, N.~R. 1985, {Solar radiophysics : studies of
  emission from the sun at metre wavelengths}

\bibitem[{{McMullin} {et~al.}(2007){McMullin}, {Waters}, {Schiebel}, {Young},
  \& {Golap}}]{mcmullin2007}
{McMullin}, J.~P., {Waters}, B., {Schiebel}, D., {Young}, W., \& {Golap}, K.
  2007, in Astronomical Society of the Pacific Conference Series, Vol. 376,
  Astronomical Data Analysis Software and Systems XVI, ed. R.~A. {Shaw},
  F.~{Hill}, \& D.~J. {Bell}, 127

\bibitem[{{Melrose}(1980)}]{melrose1980}
{Melrose}, D.~B. 1980, \ssr, 26, 3, \dodoi{10.1007/BF00212597}

\bibitem[{{Mohan}(2021)}]{mohan2021}
{Mohan}, A. 2021, \aap, 655, A77, \dodoi{10.1051/0004-6361/202142029}

\bibitem[{{Mohan} {et~al.}(2019){Mohan}, {Mondal}, {Oberoi}, \&
  {Lonsdale}}]{mohan2019}
{Mohan}, A., {Mondal}, S., {Oberoi}, D., \& {Lonsdale}, C.~J. 2019, \apj, 875,
  98, \dodoi{10.3847/1538-4357/ab0ae5}

\bibitem[{{Mondal}(2021)}]{mondal2021b}
{Mondal}, S. 2021, \solphys, 296, 131, \dodoi{10.1007/s11207-021-01877-3}

\bibitem[{Mondal {et~al.}(2019)Mondal, Mohan, Oberoi, Morgan, Benkevitch,
  Lonsdale, Crowley, \& Cairns}]{mondal2019}
Mondal, S., Mohan, A., Oberoi, D., {et~al.} 2019, The Astrophysical Journal,
  875, 97, \dodoi{10.3847/1538-4357/ab0a01}

\bibitem[{{Mondal} \& {Oberoi}(2021)}]{mondal2021a}
{Mondal}, S., \& {Oberoi}, D. 2021, \apj, 920, 11,
  \dodoi{10.3847/1538-4357/ac1076}

\bibitem[{Mondal {et~al.}(2020)Mondal, Oberoi, \& Mohan}]{mondal2020}
Mondal, S., Oberoi, D., \& Mohan, A. 2020, The Astrophysical Journal, 895, L39,
  \dodoi{10.3847/2041-8213/ab8817}

\bibitem[{{Parker}(1988)}]{parker1988}
{Parker}, E.~N. 1988, \apj, 330, 474, \dodoi{10.1086/166485}

\bibitem[{Pauluhn \& Solanki(2007)}]{pauluhn2007}
Pauluhn, A., \& Solanki, S.~K. 2007, Astronomy and Astrophysics, 462, 311,
  \dodoi{10.1051/0004-6361:20065152}

\bibitem[{{Pick} \& {Vilmer}(2008)}]{pick2008}
{Pick}, M., \& {Vilmer}, N. 2008, \aapr, 16, 1,
  \dodoi{10.1007/s00159-008-0013-x}

\bibitem[{{Reid}(2020)}]{reid2020}
{Reid}, H. A.~S. 2020, Frontiers in Astronomy and Space Sciences, 7, 56,
  \dodoi{10.3389/fspas.2020.00056}

\bibitem[{{Reid} \& {Kontar}(2017)}]{reid2017}
{Reid}, H. A.~S., \& {Kontar}, E.~P. 2017, \aap, 606, A141,
  \dodoi{10.1051/0004-6361/201730701}

\bibitem[{{Seaton} {et~al.}(2021){Seaton}, {Hughes}, {Tadikonda}, {Caspi},
  {DeForest}, {Krimchansky}, {Hurlburt}, {Seguin}, \& {Slater}}]{seaton2021}
{Seaton}, D.~B., {Hughes}, J.~M., {Tadikonda}, S.~K., {et~al.} 2021, Nature
  Astronomy, 5, 1029, \dodoi{10.1038/s41550-021-01427-8}

\bibitem[{{Sharma} {et~al.}(2022){Sharma}, {Oberoi}, {Battaglia}, \&
  {Krucker}}]{sharma2022}
{Sharma}, R., {Oberoi}, D., {Battaglia}, M., \& {Krucker}, S. 2022, \apj, 937,
  99, \dodoi{10.3847/1538-4357/ac87fc}

\bibitem[{{Sturrock}(1964)}]{sturrock1964}
{Sturrock}, P.~A. 1964, {Type III Solar Radio Bursts}, Vol.~50, 357

\bibitem[{{Tajfirouze} \& {Safari}(2012)}]{tajfirouze2012}
{Tajfirouze}, E., \& {Safari}, H. 2012, \apj, 744, 113,
  \dodoi{10.1088/0004-637X/744/2/113}

\bibitem[{{Takakura}(1967)}]{takakura1967}
{Takakura}, T. 1967, \solphys, 1, 304, \dodoi{10.1007/BF00151359}

\bibitem[{Tingay {et~al.}(2013)Tingay, Goeke, Bowman, Emrich, Ord, Mitchell,
  Morales, Booler, Crosse, Wayth, Lonsdale, Tremblay, Pallot, Colegate,
  Wicenec, Kudryavtseva, Arcus, Barnes, Bernardi, Briggs, Burns, Bunton,
  Cappallo, Corey, Deshpande, Desouza, Gaensler, Greenhill, Hall, Hazelton,
  Herne, Hewitt, Johnston-Hollitt, Kaplan, Kasper, Kincaid, Koenig,
  Kratzenberg, Lynch, McKinley, McWhirter, Morgan, Oberoi, Pathikulangara,
  Prabu, Remillard, Rogers, Roshi, Salah, Sault, Udaya-Shankar, Schlagenhaufer,
  Srivani, Stevens, Subrahmanyan, Waterson, Webster, Whitney, Williams,
  Williams, \& Wyithe}]{tingay2013}
Tingay, S.~J., Goeke, R., Bowman, J.~D., {et~al.} 2013, Publications of the
  Astronomical Society of Australia, 30, \dodoi{10.1017/pasa.2012.007}

\bibitem[{{Upendran} \& {Tripathi}(2021)}]{upendran2021}
{Upendran}, V., \& {Tripathi}, D. 2021, \apj, 916, 59,
  \dodoi{10.3847/1538-4357/abf65a}

\bibitem[{Virtanen {et~al.}(2020)Virtanen, Gommers, Oliphant, Haberland, Reddy,
  Cournapeau, Burovski, Peterson, Weckesser, Bright, {van der Walt}, Brett,
  Wilson, Millman, Mayorov, Nelson, Jones, Kern, Larson, Carey, Polat, Feng,
  Moore, {VanderPlas}, Laxalde, Perktold, Cimrman, Henriksen, Quintero, Harris,
  Archibald, Ribeiro, Pedregosa, {van Mulbregt}, \& {SciPy 1.0
  Contributors}}]{2020SciPy-NMeth}
Virtanen, P., Gommers, R., Oliphant, T.~E., {et~al.} 2020, Nature Methods, 17,
  261, \dodoi{10.1038/s41592-019-0686-2}

\bibitem[{{Wayth} {et~al.}(2018){Wayth}, {Tingay}, {Trott}, {Emrich},
  {Johnston-Hollitt}, {McKinley}, {Gaensler}, {Beardsley}, {Booler}, {Crosse},
  {Franzen}, {Horsley}, {Kaplan}, {Kenney}, {Morales}, {Pallot}, {Sleap},
  {Steele}, {Walker}, {Williams}, {Wu}, {Cairns}, {Filipovic}, {Johnston},
  {Murphy}, {Quinn}, {Staveley-Smith}, {Webster}, \& {Wyithe}}]{wayth2018}
{Wayth}, R.~B., {Tingay}, S.~J., {Trott}, C.~M., {et~al.} 2018, \pasa, 35,
  e033, \dodoi{10.1017/pasa.2018.37}

\bibitem[{{Wild} \& {Smerd}(1972)}]{wild1972}
{Wild}, J.~P., \& {Smerd}, S.~F. 1972, \araa, 10, 159,
  \dodoi{10.1146/annurev.aa.10.090172.001111}

\bibitem[{{Wild} {et~al.}(1963){Wild}, {Smerd}, \& {Weiss}}]{wild1963}
{Wild}, J.~P., {Smerd}, S.~F., \& {Weiss}, A.~A. 1963, \araa, 1, 291,
  \dodoi{10.1146/annurev.aa.01.090163.001451}

\end{thebibliography}
\bibliographystyle{aasjournal}

\end{document}